\documentclass[aip,
               reprint,
               twocolumn,
               showpacs, 
               superscriptaddress,
               preprintnumbers]{revtex4-2}
\usepackage[utf8]{inputenc}
\usepackage{graphicx}

\usepackage[dvipsnames,table]{xcolor}
\usepackage[caption=false]{subfig}
\usepackage{amsmath, amsfonts, amssymb}
\usepackage[linktocpage=true,
  colorlinks=true, 
  pdfborder={0 0 0},
  linkcolor=blue,
  citecolor=blue,
  filecolor=yellow,
  urlcolor=blue,
  bookmarks,
  pdfauthor={},
]{hyperref}
\usepackage{tikz}

\makeatletter
\newcommand*{\rom}[1]{\expandafter\romannumeral #1}
\makeatother

\begin{document}
\preprint{\textit{The following article has been submitted to Physics of Fluids.}}

\title{Linear and nonlinear active microrheology of viscous, viscoelastic, and elastic media: A fluid particle dynamics approach}
\author{Muhammed Muhsin Abdul Azeez}
\email{muhammed.abdul@ovgu.de}
\affiliation{Otto-von-Guericke-Universit\"at Magdeburg, Institut f\"ur Physik, Universit\"atsplatz 2, 39106 Magdeburg, Germany}

\author{Henning Reinken}
\email{henning.reinken@ovgu.de}
\affiliation{Otto-von-Guericke-Universit\"at Magdeburg, Institut f\"ur Physik, Universit\"atsplatz 2, 39106 Magdeburg, Germany}

\author{Andreas M. Menzel}
\email{a.menzel@ovgu.de}
\affiliation{Otto-von-Guericke-Universit\"at Magdeburg, Institut f\"ur Physik, Universit\"atsplatz 2, 39106 Magdeburg, Germany}

\newcommand{\HR}[1]{{\color{red}{#1}}}
\newcommand{\MM}[1]{{\color{blue}{#1}}}
\newcommand{\todo}[1]{{\color{magenta}{#1}}}

\date{\today}

\begin{abstract}
Active microrheology is an effective tool to determine the rheological properties of viscous, viscoelastic, or elastic materials on microscopic length scales.
The positional response of an embedded probe particle to an externally applied oscillating driving force allows to indirectly characterize the properties of the surrounding media. 
We aim to explore the linear and nonlinear response of probe particles in a microrheological setup of planar geometry. 
For this purpose, we extend the computational method of fluid particle dynamics from viscous fluid-like to viscoelastic and elastic media, including nonlinear regimes. 
We consider a system confined by solid walls. In this case, we validate the approach by quantifying the linear response in terms of a Jeffreys model.
Increasing the amplitude of the driving force, we observe distinct nonlinear effects. They include distorted stress-strain curves and a gradual net drift of probe particles initially positioned close to a wall.
This drift vanishes in the viscous fluid-like and elastic solid-like limits, but is manifest for intermediate viscoelastic systems.
We further address a setup of two probe particles in the absence of walls. They experience reciprocal pairwise oscillatory forcing. Here, nonlinearities in viscoelastic systems induce a net drift gradually moving the particles further apart from each other.
Comparing with real setups, our implementation of the driving force is in line with experimental setups of optical tweezers or active magnetic microrheology. 
\end{abstract}

\maketitle

\section{INTRODUCTION}
Viscoelastic materials exhibit both elastic energy storage and viscous dissipation under mechanical deformation~\cite{larson1999structure}. This behavior is observed in a wide range of soft matter systems with important biological and industrial applications, including polymer solutions, gels, cellular cytoplasm, and biopolymer networks~\cite{claessens2006microstructure,brust2013rheology,li2021microswimming,hemingway2015active,chaudhuri2017viscoelastic}.
Concerning biological systems, bacterial biofilms develop mechanical properties during their growth that strongly influence their mechanical stability and resistance to external stresses~\cite{flemming2010biofilm,jana2020nonlinear,hall2004bacterial}. Similarly, the mechanical properties of biological materials regulate essential physiological processes such as intracellular transport, cell migration, and mechanotransduction, where cells sense and respond to mechanical stimuli~\cite{guigas2007probing,janmey2007cell,brust2013rheology,fletcher2010cell}. Beyond biological systems, viscoelastic materials are also important in engineering applications. For example, hydrogels are widely used as tissue scaffolds~\cite{jeanie2003hydrogels}, while magnetorheological fluids and elastomers exhibit field-dependent rheological properties~\cite{carlson2000mr,bossis2002magnetorheological, han2013field, odenbach2016microstructure, schumann2017situ, kalina2023multiscale, fischer2026doubling} that enable tunable damping, vibration control, and adaptive actuation. Generally, most types of food represent another class of materials of nontrivial mechanical properties \cite{fuller2022kitchen,mathijssen2023culinary}. Characterizing these complex materials requires techniques that can probe their local mechanical properties. In that case, conventional macroscopic rheological measurements may not be sufficient.

Over the past few decades, microrheology has emerged as a powerful paradigm for probing the mechanical properties of viscoelastic materials at microscopic length scales~\cite{macintosh1999microrheology,squires2010fluid,gardel2005microrheology,waigh2005microrheology,tassieri2010measuring,fernandez2025microrheology}. Unlike conventional bulk rheology, microrheology infers the rheological response from the motion of nano- to micrometer-sized probe particles embedded within the material. While passive microrheology relies on thermal fluctuations~\cite{gittes1997microscopic,gisler1999scaling,nakayama2025microrheology,crocker2000twopoint,joyner2020microrheology,raikher2013brownian,huang2015buckling}, active microrheology employs externally applied forces to drive the probe particles~\cite{ziemann1994local,neuman2008single,wilhelm2008out,bausch1999measurement,chiang2011optical,neckernuss2016active}. Such forces are commonly applied using optical tweezers~\cite{eric2005applications,meyer2006laser} or magnetic fields~\cite{amblard1996subdiffusion,ziemann1994local,habdas2004forced}. 
In the latter case, magnetic or magnetizable particles are precisely driven via gradients of external magnetic fields.
This method has the advantage of being able to generate larger forces~\cite{waigh2005microrheology,neuman2008single}.
Thus, active microrheology enables the characterization of materials over a wider frequency range and into the nonlinear viscoelastic regime~\cite{gazuz2009active,squires2005simple,wilson2011small,khair2010active}.

Most rheological studies are performed in the linear response regime. There the deformation amplitude is sufficiently small, such that the mechanical response can be characterized by the frequency-dependent complex modulus~\cite{mason1995optical,tomaiuolo2016blood,tassieri2015linear,gomez2014probing}. Within this regime, simple linear viscoelastic models can quantify the response, such as the Maxwell and Kelvin--Voigt models and their generalizations~\cite{ferry1980viscoelastic}. 
The Maxwell model consists of an elastic element (spring) connected in series with a viscous element (dashpot), whereas in the Kelvin--Voigt model, they are connected in parallel. 
These models successfully describe the linear relaxation behavior of many viscoelastic materials~\cite{hernandez2002relaxation,nguyen2021viscoelasticity,haario2014identification,pezzinga2023charecterization}. 
However, various complex materials are subjected to large deformations, where nonlinear effects become significant, and the linear description is no longer adequate.

Capturing these nonlinear effects in computer simulations requires constitutive models that remain valid beyond the infinitesimal-strain approximation. 
At the same time, the coupling between the embedded probe particles and the surrounding viscoelastic medium must also be taken into account. 
This coupling is particularly challenging because the particle-fluid interface must satisfy the appropriate boundary conditions while the constitutive equations are solved simultaneously. 
Several numerical approaches have been developed to address this problem, including lattice-Boltzmann methods~\cite{young2017novel,min2019novel}, smoothed-particle hydrodynamics~\cite{vazquez2017sph,vazquez2012sph}, immersed-boundary methods~\cite{saadat2018immersed,shenxu2023fully,sreenath2017fully}, multiparticle collition dynamics~\cite{ji2011mesoscale}, and finite-volume or finite-element solvers~\cite{feng1996dynamic,mengfei2016numerical,jian2009fictitious,udayakumar2002interface,fernandes2022finite}. 
Although these techniques have been successfully applied to complex fluids, accurately enforcing the particle-fluid boundary conditions while retaining computational efficiency remains a significant challenge.

To this end, we adopt and adapt the fluid particle dynamics (FPD) method~\cite{tanaka2000simulation,tanaka2006viscoelastic,reinken2026hydrodynamics}. 
Particles are represented as highly viscous fluid regions embedded within the surrounding medium, which naturally satisfies the no-slip boundary condition at the particle surface. 
We here extend the method to viscoelastic and elastic media.
The viscoelastic response of the medium is described using a nonlinear neo-Hookean constitutive relation with a relaxation timescale.
To investigate the rheological properties, we embed a probe particle that is subjected to an oscillating driving force. 

First, we demonstrate that the approach reproduces the expected linear viscoelastic response in the small-deformation limit. There, the rheological properties are characterized by the complex shear modulus. Then, we investigate the nonlinear regime at larger deformation amplitudes. 
The analysis reveals higher harmonic generation in the stress-strain curves.
In particular, we include the effects of nearby walls on the motion of the probe particle.
Sufficiently close to a wall, we observe asymmetric stress-strain behavior and a gradual drift of the mean particle position away from the wall.

In an additional setup, we consider two particles experiencing equal and opposite oscillatory forcing.
Similar nonlinear effects are reproduced. Averaged over the cycles, the particles drift apart from each other in the viscoelastic regime.

Our results demonstrate that the fluid particle dynamics method can be successfully extended to viscoelastic and elastic carrier media of the immersed particles. Moreover, we show that the introduced framework is suitable to investigate nonlinear microrheological phenomena beyond the scope of conventional linear viscoelastic models.

\section{THEORETICAL DESCRIPTION AND COMPUTATIONAL METHOD}\label{sec:model}
We start by introducing the theoretical description of nonlinear viscoelasticity used in the following. To integrate elasticity and hydrodynamics, we represent all variables in Eulerian notation with spatial coordinates $\boldsymbol{r}$. In this framework, the (Eulerian) strain tensor $\boldsymbol{U}$ is given by
\begin{equation}
    \boldsymbol{U} = \frac{1}{2}\left[ \nabla \boldsymbol{u} + (\nabla \boldsymbol{u})^{\mathsf{T}} - \nabla \boldsymbol{u} \cdot (\nabla \boldsymbol{u})^{\mathsf{T}} \right],
\end{equation}
where $\boldsymbol{u}$ denotes the displacement field and $^{\mathsf{T}}$ marks the transpose. The tensor $\boldsymbol{U}$ characterizes the local deformation of the material in the current (deformed) configuration~\cite{temmen2000convective,menzel2026linear}. Time evolution of $\boldsymbol{U}$ is given by
\begin{equation}
    \partial_t \boldsymbol{U} + \boldsymbol{v} \cdot \nabla\boldsymbol{U} + (\nabla \boldsymbol{v})\cdot \boldsymbol{U} + \boldsymbol{U}\cdot (\nabla \boldsymbol{v})^{\mathsf{T}} = \boldsymbol{A}  - \frac{1}{\tau} \boldsymbol{U},
    \label{eq:U-evolution}
\end{equation}
where $\boldsymbol{A} = \frac{1}{2}\left[ \nabla \boldsymbol{v} + (\nabla \boldsymbol{v})^{\mathsf{T}} \right]$ denotes the symmetrized velocity gradient tensor. 
These dynamics for the elastic strain systematically arise in
established generalized hydrodynamics theories that include elasticity~\cite{temmen2000convective,pleiner2004nonlinear,menzel2026linear}.
Transport is included via the lower convective derivative on the left-hand side.
Generation of strain by flow gradients is given by the first term on the right-hand side.
Finally, the last term describes relaxation of elastic strain (or, through associated stress-strain relations, equivalently, of elastic stress \cite{pleiner2004nonlinear,menzel2026linear}), with $\tau$ being the associated relaxation time. A conventional illustrative picture for associated underlying microscopic processes are disentanglement dynamics of polymer chains in noncrosslinked polymer melts and solutions \cite{strobl2007physics}. 
The description includes both viscous fluid-like systems without elastic memory for $\tau \rightarrow 0$ and elastic and viscoelastic solid-like systems for $\tau \to \infty$.
Varying $\tau$ between these limiting cases encompasses a broad spectrum of viscoelastic fluids. It enables a continuous transition from viscous fluid-like via viscoelastic to elastic solid-like behavior.
For a specific viscoelastic medium described by a finite value of $\tau$, the observed properties depend on the time scale of observation. 
In particular, on timescales much smaller than $\tau$, the system behaves elastically, whereas on timescales much larger than $\tau$, it approaches a viscous fluid-like regime.

Next, we specify the constitutive relation for the stress tensor. The elastic response is assumed to be neo-Hookean.
For an incompressible neo-Hookean material, the strain energy density is given by~\cite{rivlin1948large,thomas2015compressible} 
\begin{equation}
    W\left(I_1\right) = \frac{\mu}{2}\left( I_1 - d \right),
    \label{eq:energy-density}
\end{equation}
where $I_1 = \mathrm{tr}(\boldsymbol{B})$ is the first invariant of the left Cauchy–Green strain tensor $\boldsymbol{B}$, and $d$ is the spatial dimension. The tensor $\boldsymbol{B}$ is related to the Eulerian strain tensor $\boldsymbol{U}$ via
\begin{equation}
    \boldsymbol{B} = (\boldsymbol{I} - 2\boldsymbol{U})^{-1},
    \label{eq:CG-tensor}
\end{equation}
where $\boldsymbol{I}$ is the unit matrix.
The parameter $\mu$ in Eq.~\eqref{eq:energy-density} is the shear modulus in the limit of small strain. 
For such a material, the
Cauchy stress tensor $\boldsymbol{\sigma}$ is given by
\begin{equation}
\boldsymbol{\sigma} ={} - p \boldsymbol{I} + \mu \boldsymbol{B} + 2 \eta \boldsymbol{A},
\label{eq:cauchy-stress}
\end{equation}
where $p$ is the pressure.
The last term describes viscous dissipation with viscosity $\eta$, which originates from internal friction during gradient flow.
The viscoelastic medium is treated as an incompressible material. Its dynamics are governed by the conservation of mass and momentum. Accordingly, the flow field $\boldsymbol{v}(\boldsymbol{r}, t)$ is assumed to respect the condition
\begin{equation}\label{eq:incomp}
    \nabla\cdot \boldsymbol{v} = 0,
\end{equation}
which follows from the continuity equation for incompressible systems. In this case, the Navier--Stokes equation becomes
\begin{equation}
\rho \partial_t \boldsymbol{v} + \rho \boldsymbol{v}\cdot \nabla \boldsymbol{v} = \nabla \cdot \boldsymbol{\sigma} + \boldsymbol{f}_\mathrm{p} + \boldsymbol{f}_\mathrm{w},
\label{eq:navier-stokes}
\end{equation}
where $\rho$ denotes the mass density. 
The pressure contained within the Cauchy stress tensor, see Eq.~(\ref{eq:cauchy-stress}), here acts as a Lagrange multiplier enforcing incompressibility according to Eq.~(\ref{eq:incomp}). The force density $\boldsymbol{f}_\mathrm{p}$ arises from forces acting on the suspended micron-sized particles, see below. 
We further include rigid, immovable walls demarcating the boundaries of our system. Their influence on the dynamics is included via the additional force density $\boldsymbol{f}_\mathrm{w}$.
This method ensures the immovability of the walls via strong linear frictional damping~\cite{goldstein1993modeling,kevlahan2001computational}.
To this end, an additional phase field is introduced, which characterizes the area of the finitely extended walls at the system boundaries. 
Here, we incorporate an additional linear friction term that strongly damps the velocity field and effectively ensures that it vanishes within the walls. 
We refer to Appendix~\ref{App:rescaling} for our rescaling of the equations and to Appendix~\ref{App:numerical-methods} for more details on the implementation of the simulations, including the rigid walls.

To investigate the microrheological response of the viscoelastic medium, we introduce an embedded particle and probe the system using an oscillatory forcing.
Numerically describing such a system is challenging, as it requires coupling the discrete particle dynamics with the macroscopic continuum description of the surrounding medium. To this end, we assume no-slip boundary conditions at the interface between the particle and the surrounding medium.
We then employ the fluid particle dynamics (FPD) method~\cite{tanaka2000simulation,tanaka2006viscoelastic, reinken2026hydrodynamics}. 
Within this approach, particles are incorporated into the continuum description and treated as fluid-like regions with significantly higher shear viscosity than the surrounding medium. 
This enhanced viscosity hinders deformation.
Specifically, we extend the FPD method to the case of viscoelastic and elastic systems.
Analogously to the shear viscosity, we increase the shear modulus within the regions covered by particles.

The $\eta_0$ and $\mu_0$ denote shear viscosity and shear elastic modulus of the viscoelastic medium, respectively. Corresponding values within regions covered by particles are parameterized as $\eta_\mathrm{p} = R_\eta \eta_0$ and $\mu_\mathrm{p} = R_\mu \mu_0$, with $R_\eta, R_\mu \gg 1$. The presence of a particle centered at position $\boldsymbol{R}(t)$ is represented by a continuous phase field $\phi(\boldsymbol{r},t)$, 
\begin{equation}
    \phi(\boldsymbol{r},t) = \frac{1}{2}\left[\tanh{\left( \frac{a - |\boldsymbol{r} - \boldsymbol{R}(t)|}{c} \right)} + 1\right],
    \label{eq:phi-filed}
\end{equation}
where $a$ denotes the radius of the particle and $c$ characterizes the thickness of the interface between the particle and the surrounding medium in the FDM approach. Consequently, the material parameters $\mu$ and $\eta$ appearing in Eq.~\eqref{eq:cauchy-stress} become spatially dependent and take the form
\begin{align}
    \eta(\boldsymbol{r}) &= \eta_0 + (R_\eta - 1)\eta_0\phi(\boldsymbol{r}), \\
    \mu(\boldsymbol{r}) &= \mu_0 + (R_\mu - 1)\mu_0\phi(\boldsymbol{r}). 
\end{align}

The particle is subjected to an oscillating force
\begin{equation}
\boldsymbol{F}(t) = F_0 \cos(\omega t) \hat{\boldsymbol{e}}_x.
\label{eq:drivingforce}
\end{equation}
$F_0$ and $\omega$ denote the amplitude and frequency of the driving force, respectively, and $\hat{\boldsymbol{e}}_x$ is the unit vector along the $x$-direction. 
The force density $\boldsymbol{f}_\mathrm{p}$ in Eq.~\eqref{eq:navier-stokes} is determined from the force $\boldsymbol{F}$ via
\begin{equation}
    \boldsymbol{f}_\mathrm{p}(\boldsymbol{r},t) = \frac{1}{V_\mathrm{p}}\phi(\boldsymbol{r},t) \boldsymbol{F}(t) .
    \label{eq:force-density}
\end{equation}
$V_\mathrm{p}$ is the volume of the particle in three dimensions or its area in two dimensions.
In this way, forces $\boldsymbol{F}(t)$ that act on a particle are distributed over the whole body of the particle.
The resulting force density then enters the Navier--Stokes equation, Eq.~(\ref{eq:navier-stokes}), and acts on the continuous medium, possibly leading to its motion.

\begin{figure}
    \centering
    \includegraphics[width=\linewidth]{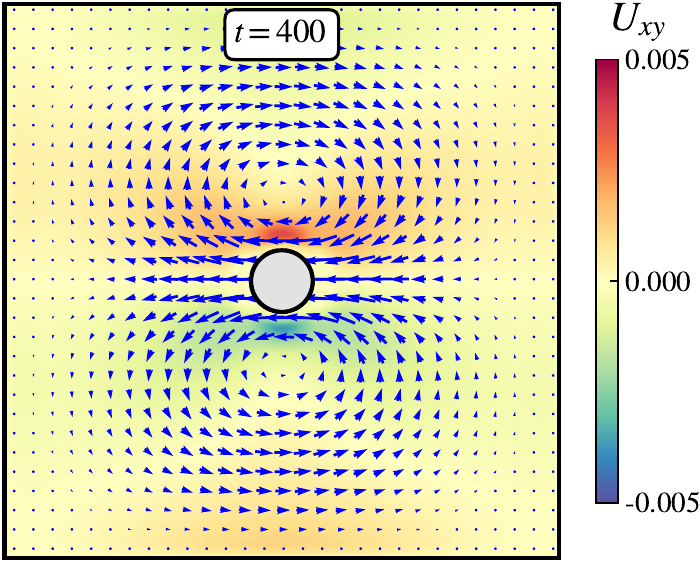}
    \caption{Snapshot of the simulation at time $t = 400$. The domain size is $18a\times 18a$, with $a$ the radius of the particle. Blue arrows indicate the velocity field, and the colormap quantifies the components $U_{xy}$ of the strain tensor $\boldsymbol{U}$. Other parameters in this simulation in rescaled units are set to $\tau = 100$ for the viscoelastic relaxation time, $\mu_0 = 0.05$ for the elastic shear modulus of the embedding medium, $F_0 = 0.01$ for the amplitude of the oscillatory driving force of the particle, and $\omega = 0.002\pi$ for its frequency.}
    \label{fig:snapshot}
\end{figure}

For numerical implementation, we rescale the relevant physical quantities into dimensionless form. The radius $a$ of the particle is chosen as the characteristic length scale, and the inertial time scale $\rho a^2/\eta_0$ is used as the characteristic time scale.
A detailed description of rescaling is given in Appendix~\ref{App:rescaling}.
In rescaled units, the shear modulus is given by $\tilde{\mu} = {\rho a^2 \mu}/{\eta_0^2}$.
From now on, we work in dimensionless units and omit the tilde on rescaled quantities.

In this initial contribution, we perform our simulations in two dimensions. Our system consists of a square domain of size $L\times L$. The thickness of the interface $c$ in Eq.~\eqref{eq:phi-filed} is set to $c = 0.2a$, and the scaling factors $R_\eta$ and $R_\mu$ are chosen as $R_\eta = R_\mu = 50$. 
This value represents a tradeoff between the accuracy of the simulations and numerical efficiency~\cite{tanaka2000simulation}.
Numerical integration of Eqs.~\eqref{eq:U-evolution} and \eqref{eq:navier-stokes} are performed using an implicit Euler scheme with a time step $\Delta t = 0.1$. The particle velocities $\mathbf{V}(t)$ are obtained by integrating the velocity field $\boldsymbol{v}(\boldsymbol{r},t)$ over the particle domains, weighted by the phase field. Integrating $\mathbf{V}(t)$ in time provides the associated position of the particle $\mathbf{R}(t)$. More details on the simulation method are reported in Appendix~\ref{App:numerical-methods}.

\begin{figure*}
\centering
\includegraphics[width=0.99\linewidth]{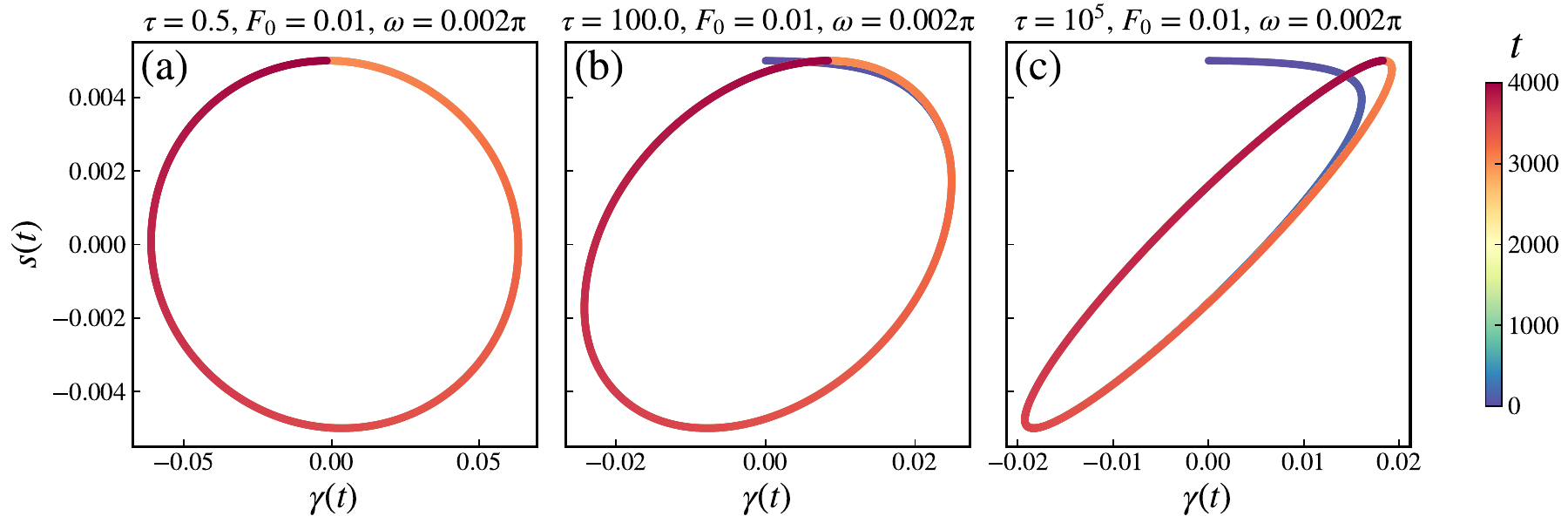}
\caption{Microrheological stress-strain curves, that is, effective stress $s(t)$, related to the force acting on the particle, as a function of effective strain $\gamma(t)$, related to the displacement of the particle, in the linear regime. With increasing value of relaxation time $\tau$, the degree of elasticity in the system increases from (a) still relatively viscous, $\tau = 0.5$, via (b) notably viscoelastic, $\tau = 100$, to (c) highly elastic, $\tau = 10^5$.
The curves cover the first four oscillation cycles, with the color scale indicating time $t$.
Other fixed parameters are the rescaled shear elastic modulus of the background material $\mu_0 = 0.02$, amplitude of the applied oscillating force $F_0 = 0.01$, and its oscillation frequency $\omega = 0.002\pi$.}
\label{fig:stress-strain-linear}
\end{figure*}

\section{RESULTS AND DISCUSSION}\label{sec:result}
We start by investigating a configuration in which the probe particle is inserted at the center of the system. It is then driven by the oscillatory force according to Eq.~(\ref{eq:drivingforce}).
A snapshot of the simulation setup is shown in Fig.~\ref{fig:snapshot}. 
The primary quantity of interest is the linear strain $\gamma(t)$. We define it as the distance the particle has moved from its initial position $\mathbf{R}(0)$ at time $t$, that is,
\begin{equation}
   \gamma(t) = \frac{1}{L_\mathrm{eff}}{\big[R_x(t)- R_x(0)\big]}, \label{eq:linear-strain}
\end{equation}
where $R_x$ is the $x$ position of the particle.
The length $L_{\mathrm{eff}}$ is an effective length scale that we here take as the diameter of the particle. 
The corresponding stress $s(t)$ is directly proportional to the force applied to the particle and is denoted as
\begin{equation}
    s(t) =  \frac{F(t)}{L_{\mathrm{eff}}^{d-1}},
    \label{eq:linear-stress}
\end{equation}
where $d$ is the spatial dimension.

\subsection{Linear microrheological response}
We first analyze the linear viscoelastic regime. 
In Fig.~\ref{fig:stress-strain-linear}, we plot the stress-strain curve for different values of $\tau$. To ensure that nonlinearities remain low in effect on the response, the amplitude of the driving force is set to $F_0 = 0.01$. Similarly, the driving frequency is chosen as $\omega = 0.002\pi$ to ensure that inertial effects remain negligible. This value of the driving frequency is used throughout. In the limit $\tau\to 0$, our parameter settings corresponds to an observed maximum Reynolds number of $\mathrm{Re}\approx 10^{-3}$. Here, the Reynolds number is defined as $\mathrm{Re} = L_\mathrm{eff}\rho V/\eta_0$, where $V$ is the velocity of the particle.

When the relaxation time $\tau$ is very small, the elastic contribution relaxes rapidly. Thus, the material exhibits predominantly viscous behavior. 
Consequently, the stress is approximately proportional to the shear rate. The stress-strain curve takes the form of a circle, as shown in Fig.~\ref{fig:stress-strain-linear}(a). Over four consecutive oscillation cycles, the curves overlap completely, indicating that the system has reached a steady state. 

For larger relaxation times, the elastic contribution becomes increasingly important. When $\tau = 100$, the stress relaxation time is comparable to the oscillation period, resulting in a finite phase difference between stress and strain. The stress-strain curve becomes of elliptic shape, as shown in Fig.~\ref{fig:stress-strain-linear}(b). 

As $\tau$ is increased further, the stress becomes increasingly in phase with the strain. However, due to the presence of the background viscosity, a finite viscous contribution remains, preventing the stress-strain curve from collapsing into a straight line. Consequently, the curve becomes a highly elongated ellipse, approaching the elastic limit, as shown in Fig.~\ref{fig:stress-strain-linear}(c).

\begin{figure}
    \centering
    \includegraphics[width=0.9\linewidth]{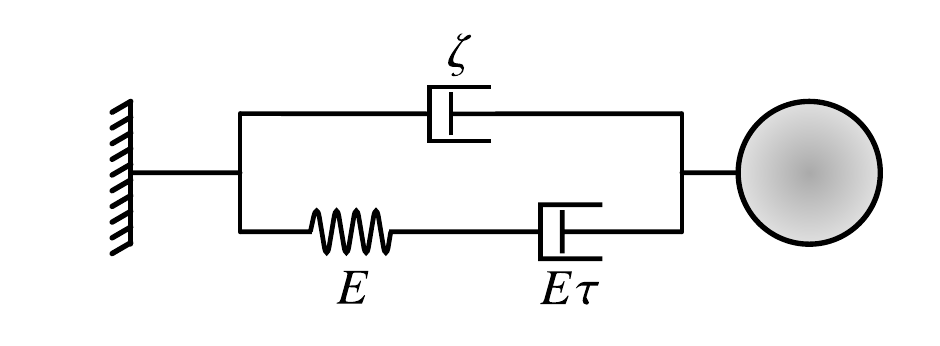}
    \caption{Schematic of the three-parameter viscoelastic model corresponding to Eq.~\eqref{eq:linear-model}. The model consists of a Maxwell element connected in parallel with a dashpot of viscosity $\zeta$. The Maxwell element (bottom) comprises a spring of modulus $E$ connected in series with a dashpot of viscosity $E\tau$. This model is commonly referred to as the Jeffreys model~\cite{khadrawi2005basic}.
    }
    \label{fig:linear-model}
\end{figure}

The elastic and viscous contributions to the material response are related to the storage modulus $G'$ and loss modulus $G''$, respectively. These correspond to the real and imaginary parts of the complex shear modulus $G^*$. In the overdamped limit, where inertial effects are negligible, $G^*$ is defined as
\begin{equation}
    G^*(\omega) = \frac{\mathcal{F}\left\{ s(t) \right\}(\omega)}{\mathcal{F}\left\{ \gamma(t) \right\}(\omega)}.
    \label{eq:complex-modulus}
\end{equation}
$\mathcal{F}\{\cdots\}$ denotes the Fourier transform. 
In the limit of small-amplitude oscillatory forcing, $\boldsymbol{B}$ in Eq.~\eqref{eq:CG-tensor} becomes 
\begin{equation}
    \boldsymbol{B} \approx \boldsymbol{I} + 2\boldsymbol{U}.
    \label{eq:linear-CG-tensor}
\end{equation}
Accordingly, the evolution equation for $\boldsymbol{U}$ becomes
\begin{equation}
    \dot{\boldsymbol{U}} = \boldsymbol{A} - \frac{1}{\tau} \boldsymbol{U}.
    \label{eq:U-evolution-linear}
\end{equation}
Combining Eqs.~\eqref{eq:cauchy-stress} and \eqref{eq:linear-CG-tensor}, then only considering the deviatoric (traceless) part of the Cauchy stress, $\boldsymbol{\sigma}^\mathrm{dev}$, yields in rescaled units
\begin{equation}
\boldsymbol{\sigma}^\mathrm{dev} = 2 \mu_0 \boldsymbol{U} + 2 \boldsymbol{A}.
\label{eq:deviatoricStressLinear}
\end{equation}
Eliminating $\boldsymbol{U}$ using Eq.~\eqref{eq:U-evolution-linear}, we obtain an evolution equation for $\boldsymbol{\sigma}^\mathrm{dev}$,
\begin{equation}
\dot{\boldsymbol{\sigma}}^\mathrm{dev} + \frac{\boldsymbol{\sigma}^\mathrm{dev}}{\tau} = 2 \left( \mu_0 + \frac{1}{\tau} \right) \boldsymbol{A} + 2 \dot{\boldsymbol{A}}.
\label{eq:evolutionDeviatoricStressLinear}
\end{equation}
Equation~(\ref{eq:evolutionDeviatoricStressLinear}) has the same general form as the differential equation describing the relation between stress $s(t)$ and strain $\gamma(t)$ in the three-parameter linear viscoelastic Jeffreys model~\cite{joseph1990fluid,khadrawi2005basic},
\begin{equation}
    \dot{s}(t) + \frac{s(t)}{\tau} = \left( E + \frac{\zeta}{\tau}\right) \dot{\gamma}(t) + \zeta \ddot{\gamma}(t).
    \label{eq:linear-model}
\end{equation}
Here, $\dot{\gamma}$ takes the role of $\boldsymbol{A}$, while $s$ takes the role of $\boldsymbol{\sigma}^\mathrm{dev}$.
The coefficients $E = \alpha \mu_0$ and 
$\zeta = \alpha $ 
denote the effective elastic modulus and viscosity, respectively. Moreover, the parameter $\alpha$ is a geometric factor relating the effective parameters $E$ and $\zeta$ in the linear model for the motion of the particle to the rheological parameters of the entire system, including the finite size and shape of the particle as well as the role of the surrounding material, its shear modulus and viscosity.
Figure~\ref{fig:linear-model} shows an illustration of the Jeffreys model. 

Taking the Fourier transform of Eq.~\eqref{eq:linear-model} and substituting into Eq.~\eqref{eq:complex-modulus}, we get
\begin{equation}
    G^*(\omega) = i\omega \zeta + \frac{i \omega E }{i \omega + {1}/{\tau}}.
\end{equation}
Thus, the expressions for $G'$ and $G''$ are given by
\begin{align}
    G'(\omega) &= \frac{E \omega^2\tau^2}{\omega^2\tau^2 + 1}, \label{eq:storage-modulus} \\
    G''(\omega) &= \zeta\omega +  \frac{E \omega \tau}{\omega^2\tau^2 + 1}. \label{eq:loss-modulus}
\end{align}

\begin{figure}
\centering
\includegraphics[width=\linewidth]{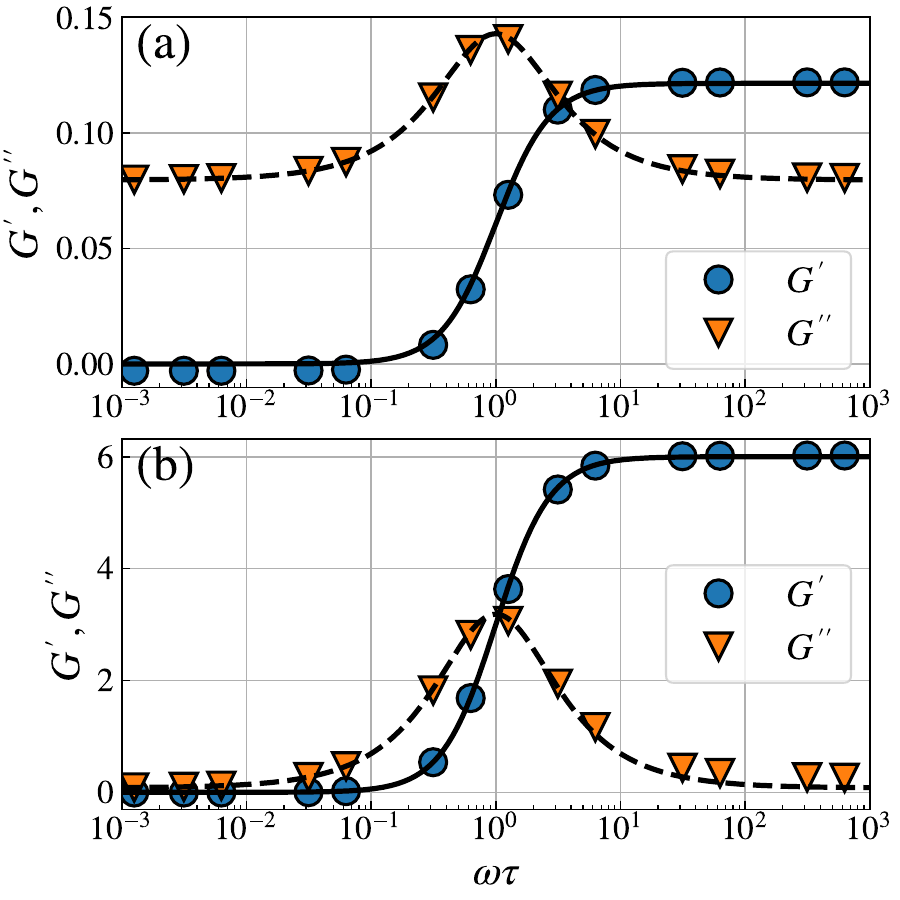}
\caption{Storage modulus $G'$ and loss modulus $G''$ in rescaled units for our microrheological setup evaluated in the linear regime are plotted as functions of the Deborah number $\omega \tau$ for rescaled elastic shear modulus (a) $\mu_0 = 0.01$ and (b) $\mu_0 = 0.5$. We vary the Deborah number by fixing the frequency of the driving force at $\omega = 0.002\pi$ and increasing $\tau$, which changes the nature of the background medium from viscous fluid-like, via viscoelastic, towards elastic. Solid and dashed lines represent the fits to $G'$ and $G''$ respectively, using Eqs.~\eqref{eq:storage-modulus} and \eqref{eq:loss-modulus}.
Our only fit parameter is the geometric prefactor, see the definition below Eq.~\eqref{eq:linear-model}, for which we find $\alpha \approx 12.3$.
The rescaled amplitude of the driving force is fixed at $F_0 = 0.01$.}
\label{fig:linear-moduli}
\end{figure}

In Fig.~\ref{fig:linear-moduli}, we plot the storage and loss moduli as a function of Deborah number, $\mathrm{De} = \omega \tau$, for two different values of the shear modulus $\mu_0$. 
We obtain the curves by varying $\tau$, while keeping $\omega$ constant.
When $\omega \tau$ is small, the relaxation time is short compared to the oscillation period, and the medium exhibits predominantly viscous fluid-like behavior. As $\omega \tau$ increases, elastic effects become increasingly important, and the medium exhibits progressively more viscoelastic and solid-like behavior.  The solid and dashed lines in Fig.~\ref{fig:linear-moduli} correspond to fits to Eqs.~\eqref{eq:storage-modulus} and \eqref{eq:loss-modulus}, respectively. The fitting parameter is found to be $\alpha \approx 12.3$.
We observe that the simulation results are in excellent agreement with the predictions of the linear model.

\begin{figure*}[!ht]
    \centering
    \includegraphics[width=0.99\linewidth]{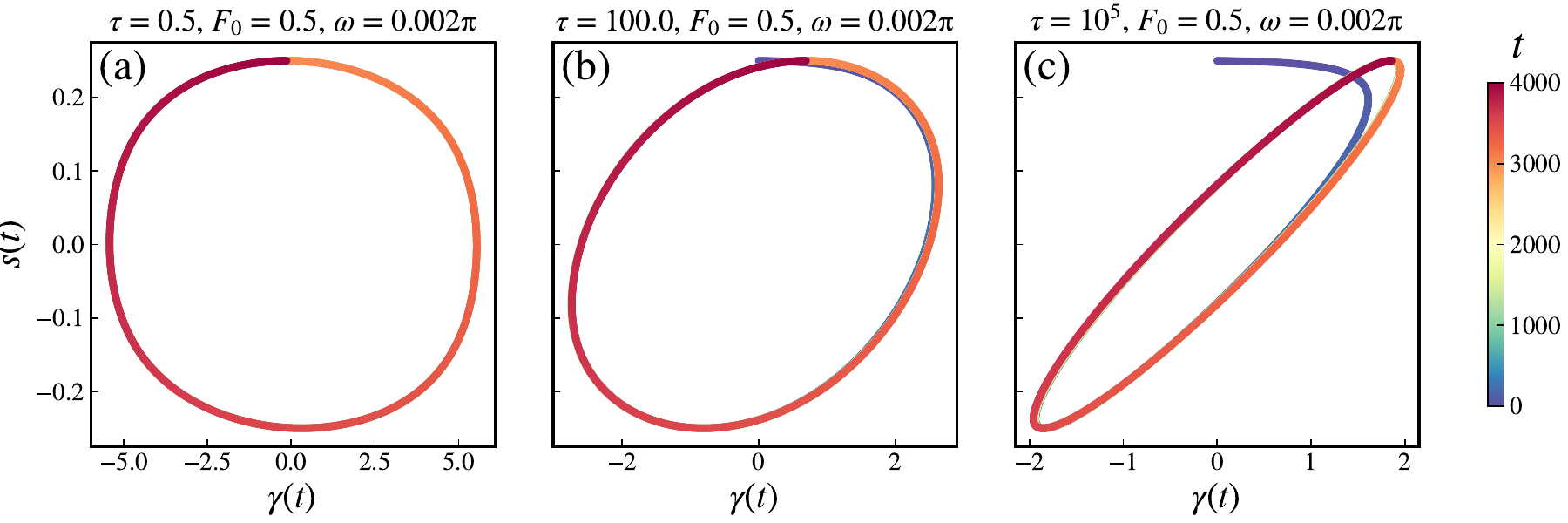}
    \caption{Microrheological stress-strain curves displayed in a similar way as in Fig.~\ref{fig:stress-strain-linear}, but now in the nonlinear regime, for different values of the relaxation time (a) $\tau = 0.5$, (b) $\tau = 100$, and (c) $\tau = 10^5$. Compared to the situation in Fig.~\ref{fig:stress-strain-linear}, the amplitude of the oscillating driving force was set to an elevated value of $F_0 = 0.5$ in rescaled units. Initially, the probe particle is positioned into the center of the computational setup. 
    The curves cover the first four oscillation cycles, with the color scale indicating rescaled time $t$. Deviations from the linear regime are specifically visible when comparing panels (a) of this figure and Fig.~\ref{fig:stress-strain-linear}. 
    Fixed remaining rescaled parameters are $\mu_0 = 0.02$ and $\omega = 0.002\pi$.}
    \label{fig:stress-strain-nonlinear_center}
\end{figure*}
When $\omega \tau \ll 1$, it can be seen that $G' \approx 0$ and $G'' \approx \zeta \omega$. In this limit, the relaxation time is much shorter than the oscillation period, so the Maxwell element (containing the spring) in Fig.~\ref{fig:linear-model} 
contributes negligibly to the response. 
Consequently, the medium behaves similarly to a viscous Newtonian fluid, with only the dashpot of viscosity $\zeta$ contributing. 

When $\omega \tau \gg 1$, the stress relaxes negligibly over a single oscillation cycle, causing the Maxwell element to behave as a purely elastic spring. Consequently, the storage modulus approaches $G' = E$. Due to the background viscosity, the loss modulus in this limit is given by $G'' = \zeta \omega$. 
This behavior is particularly evident in Fig.~\ref{fig:linear-moduli}(a) for $\mu_0 = 0.01$. 

In the limits $\omega \tau \to 0$ and $\omega \tau \to \infty$, $G''$ approaches a finite value, which quantifies the background viscous dissipation. For intermediate values of $\omega \tau$, viscoelasticity contributes in addition.
This results in a nonmonotonic dependence of $G''$ on $\omega \tau$. The loss modulus attains a maximum of $G'' = \zeta \omega + E/2$ at $\omega \tau = 1$. When the value of $\mu_0$ is increased, see Fig.~\ref{fig:linear-moduli}(b), the role of viscoelasticity increases relative to background viscous dissipation. Thus, the maximum of $G''$ shifts to larger values. 

Contrarily to the loss modulus, the storage modulus $G'$ exhibits a monotonic increase as a function of $\omega\tau$. It transitions from its minimum value, $G' = 0$, at low $\omega \tau$ to its maximum value, $G' = E$, as $\omega \tau$ increases.  
The halfway value $E/2$ is found at $\omega \tau = 1$.

\begin{figure}
    \centering
    \includegraphics[width=\linewidth]{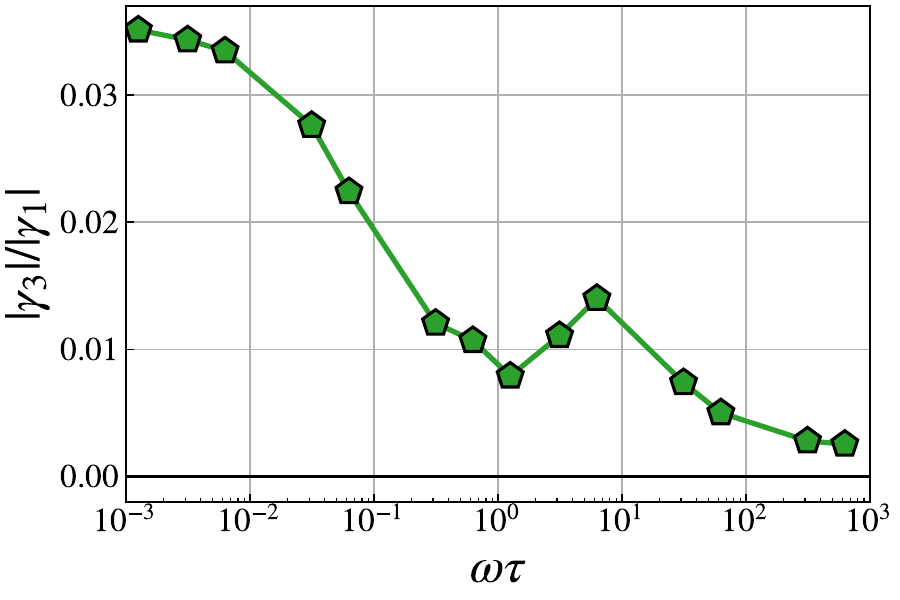}
    \caption{Amplitude of the third harmonic coefficient of the strain $|\gamma_3|$, see Eq.~\eqref{eq:fourier-series}, relative to that of the fundamental response $|\gamma_1|$ in the nonlinear regime of the microrheological setup considered in Fig.~\ref{fig:stress-strain-nonlinear_center} as a function of the Deborah number $\omega \tau$. The probe particle is initially positioned at the center of the simulation setup. We show averages over the second and third cycle. While an overall decreasing trend emerges, a local minimum appears at $\omega\tau\approx1$, where in Fig.~\ref{fig:linear-moduli} the maximum of $G''$ is located in the linear case. The fixed remaining rescaled parameters are again $F_0 = 0.5$, $\mu_0 = 0.02$, and $\omega = 0.002\pi$.}
    \label{fig:gamma-n_center}
\end{figure}

\subsection{Nonlinear microrheological response}

So far we have restricted our analysis to the linear viscoelastic and linear elastic regime. There, the deformation amplitude was sufficiently small, so that it was justified to neglect nonlinear contributions. However, one of the key advantages of our approach is its ability to capture nonlinear viscoelastic and elastic effects arising at larger deformations.
Still, inertial effects remain negligible. 
To investigate this nonlinear regime, we increase the amplitude of the driving force in rescaled units to $F_0 = 0.5$. We analyze the resulting stress-strain behavior. 
For our analysis, we set $\mu_0 = 0.02$, intermediate to the values considered for the linear regime in Fig.~\ref{fig:linear-moduli}.

Figure~\ref{fig:stress-strain-nonlinear_center} shows the stress-strain curves for different values of the relaxation time.
Here, Fig.~\ref{fig:stress-strain-nonlinear_center}(a) illustrates the almost purely viscous fluid-like situation, Fig.~\ref{fig:stress-strain-nonlinear_center}(b) shows an exemplary situation of a notably viscoelastic fluid, and Fig.~\ref{fig:stress-strain-nonlinear_center}(c) contains a relatively solid-like viscoelastic case. 
Compared to the linear regime shown in Fig.~\ref{fig:stress-strain-linear}, the stress-strain curves exhibit clear distortions, indicating the increasing influence of nonlinear behavior.
In particular, the curve in the viscous case in Fig.~\ref{fig:stress-strain-nonlinear_center}(a) shows deviations from the circular curve in the linear case, see Fig.~\ref{fig:stress-strain-linear}(a).
Similar effects are still visible in the viscoelastic regime at $\tau = 100$ in Fig.~\ref{fig:stress-strain-nonlinear_center}(b).
However, the stress-strain curve in the solid-like case, illustrated in Fig.~\ref{fig:stress-strain-nonlinear_center}(c), still resembles the one in the linear case.

\begin{figure*}
    \centering
    \includegraphics[width=0.99\linewidth]{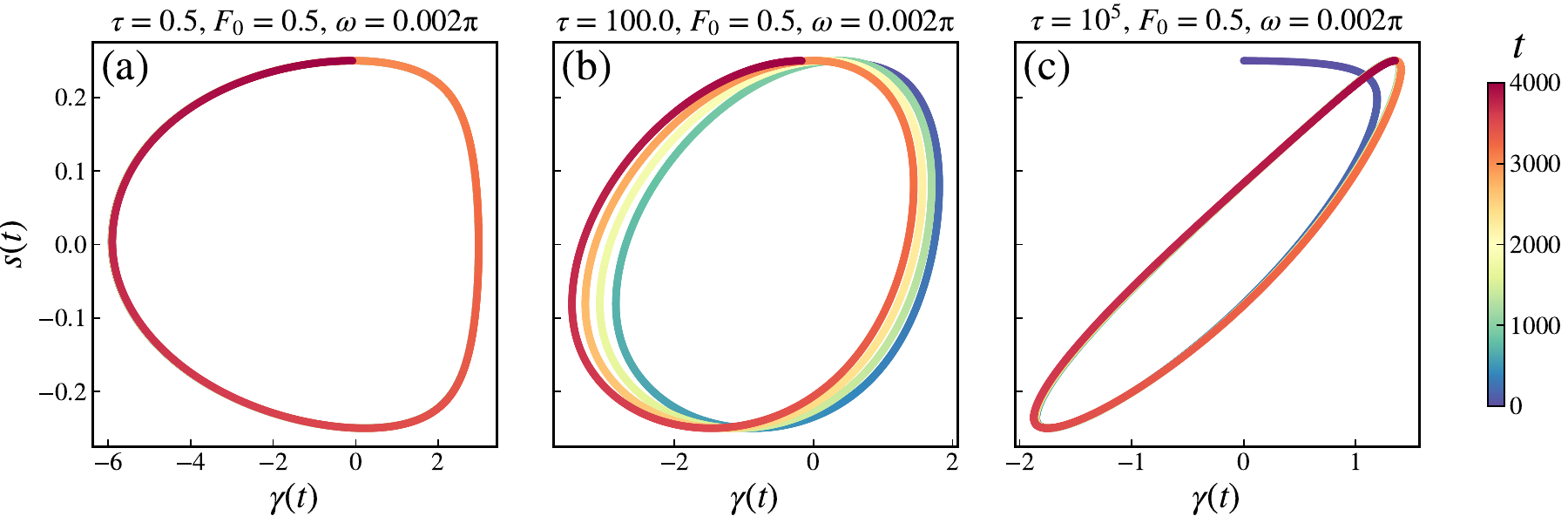}
    \caption{Microrheological stress-strain curves in the nonlinear regime, for different values of the relaxation time (a) $\tau = 0.5$, (b) $\tau = 100$, and (c) $\tau = 10^5$. The simulation setup here is identical to the one in Fig.~\ref{fig:stress-strain-nonlinear_center}. However, the probe particle here is initially positioned asymmetrically, that is, off-center, into the computational setup, at a distance of $4a$ to the right of the center. 
    The curves cover the first four oscillation cycles.
    Fixed remaining rescaled parameters are again $\mu_0 = 0.02$ and $\omega = 0.002\pi$.}
    \label{fig:stress-strain-nonlinear_off-center}
\end{figure*}

\begin{figure}
    \centering
    \includegraphics[width=\linewidth]{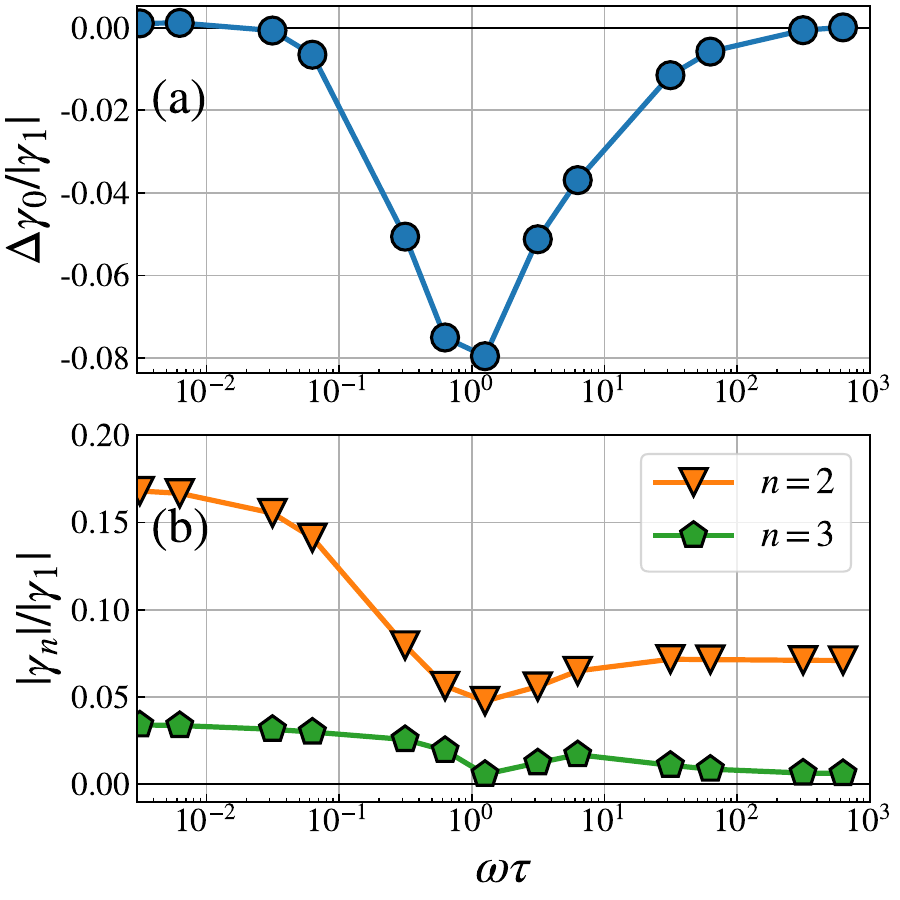}
    \caption{Coefficients of the zeroth, second, and third harmonics of the induced strain, see Eq.~\eqref{eq:fourier-series}, in the nonlinear regime as a function of the Deborah number $\omega\tau$ for the microrheological setup considered in Fig.~\ref{fig:stress-strain-nonlinear_off-center}. Here, the probe particle is initially positioned by a distance $4a$ off the center of the computational setup. (a) The relative difference $\Delta\gamma_0/|\gamma_1|$ in the values of the zero mode between the second and third cycle quantifies the associated drift in position away from the closer wall. (b) A relative amplitude of the second harmonic $|\gamma_2|/|\gamma_1|$ further indicates the asymmetry in the underlying placement of the particle within the microrheological setup. Conversely, the relative amplitude of the third harmonic $|\gamma_3|/|\gamma_1|$ shows a trend similar to the symmetric situation in Fig.~\ref{fig:gamma-n_center}. Here, we take averages over the second and third cycle. Remaining fixed rescaled parameter values are $F_0 = 0.5$, $\mu_0 = 0.02$, and $\omega = 0.002\pi$.}
    \label{fig:gamma-n_off-center}
\end{figure}

To quantify these observations, we decompose the response of the material into harmonics. 
In particular, the nonlinearity manifests itself through the generation of higher-order contributions. 
The applied stress is given by
\begin{equation}
    s(t) = s_1 e^{i\omega t} + \mathrm{c.c.},
\end{equation}
where $\mathrm{c.c}$ denotes the complex conjugate.
In the linear regime, the response remains confined to this single oscillation frequency $\omega$ and can be written as 
$\gamma(t) = \gamma_1 \exp(i\omega t ) + \mathrm{c.c.}$.
However, this simple relation no longer holds in the nonlinear regime. There, the resulting response, given by induced strain, develops contributions of higher harmonics. In general, this response can be expressed as a Fourier series
\begin{equation}\label{eq:fourier-series}
    \gamma(t) = \sum_{n=-\infty}^{\infty}\gamma_n e^{in\omega t},
\end{equation}
where $\gamma_n$ are the complex Fourier coefficients corresponding to the $n$-th harmonics. 
Since $\gamma(t)$ is real, $\gamma_{-n}$ are the complex conjugates of $\gamma_n$. 
The coefficient $\gamma_{1}$ corresponds to the fundamental response, which is the only nonzero component in the linear case.
All coefficients $\gamma_{|n|>1}$ describe contributions by higher harmonics.

Here, we initially position the particle in the center of the computational setup. Then, we focus on the third component $\gamma_3$.
Figure~\ref{fig:gamma-n_center} shows $\gamma_3$ normalized by the fundamental component $\gamma_1$ as a function of the Deborah number $\omega \tau$.
It shows a decreasing trend from high values in the viscous regime to almost zero towards the elastic regime.
Thus, overall nonlinear effects become less significant as $\omega \tau$ increases. 
This is in line with the force-controlled protocol we apply.
While varying the relaxation time $\tau$, we keep amplitude and oscillation frequency of the driving force constant.
As a result, the maximally reached strain is smaller when elastic effects increasingly come into play, see Fig.~\ref{fig:stress-strain-nonlinear_center}, decreasing the relative significance of nonlinear effects.

Moreover, we note that Fig.~\ref{fig:gamma-n_center} displays a local minimum of $|\gamma_3|/|\gamma_1|$ at $\omega \tau \approx 1$. Here, the inverse of the oscillation frequency of the externally applied driving force approximately matches the viscoelastic relaxation time.
In the linear case, this match leads to the maximum in the loss modulus, see Fig.~\ref{fig:linear-moduli}.
As our results show, this matching of time scales still leads to a characteristic signature in the nonlinear response for the depicted higher harmonic.

\subsection{Nonsymmetric geometry in the nonlinear regime}

We return to the Fourier expansion of the strain $\gamma(t)$, see Eq.~\eqref{eq:fourier-series}, in response to the imposed stress. There, the overall, time-averaged mean is given by $\gamma_0$. It must vanish therefore in a symmetric situation. 
Moreover, all even higher harmonics must vanish in a symmetric situation, because there is an exact antisymmetric correspondence between the displacement induced by forcing into opposite directions. In our computational setup, we expect that such a symmetric behavior can be observed approximately, if the particle is initially positioned at the center of the simulation box. This is indeed the case. 

However, if the particle is initially positioned off-center, that is, closer to one of the walls, the situation changes. 
In this case, even harmonics can be generated in the response.
They result from the asymmetric influence of the presence of the walls, which can then be inferred from the response of the particle. 

Corresponding stress-strain curves for different values of the relaxation times are included in Fig.~\ref{fig:stress-strain-nonlinear_off-center}.
Here, the particle is initially displaced by $4a$ from the center of the system to the right.
Again, Fig.~\ref{fig:stress-strain-nonlinear_off-center}(a) shows a viscous fluid-like situation, Fig.~\ref{fig:stress-strain-nonlinear_off-center}(b) the situation of a viscoelastic fluid, and Fig.~\ref{fig:stress-strain-nonlinear_off-center}(c) the viscoelastic solid-like case. 
It is evident from Fig.~\ref{fig:stress-strain-nonlinear_off-center}(b) that the stress-strain curves between successive oscillation cycles do not overlap, unlike Fig.~\ref{fig:stress-strain-nonlinear_center}, where the curves for different cycles approximately coincide. 

To quantify this net drift of the particle, we define $\Delta\gamma_0$ as the difference in $\gamma_0$ between successive oscillation cycles. $\gamma_0$ is calculated as the average of $\gamma(t)$ over one cycle.
In Fig.~\ref{fig:gamma-n_off-center}(a), we plot $\Delta\gamma_0/|\gamma_1|$ as a function of $\omega\tau$. 
Initially, in the viscous fluid-like regime ($\omega \tau \ll 1$), $\Delta\gamma_0/|\gamma_1|$ is close to zero.
This is consistent with the restrictions of viscous Stokes flow, where time-reversal symmetry holds.
Thus, the particle retraces the same trajectory when the direction of the oscillating force reverses. Thus, there is no net cycle-averaged drift.

The situation changes in the viscoelastic regime, where memory effects are present.
Here, the stress depends on the history of deformation.
As Fig.~\ref{fig:gamma-n_off-center}(a) shows, the result is a net drift away from the closer wall.
The maximum effect is observed when the inverse of the oscillation frequency matches the viscoelastic relaxation time, $\omega \tau \approx 1$.

Approaching the elastic regime for $\omega \tau \gg 1$, the net drift vanishes again, consistently with a solid-like material. There, net flow is impossible.
The particle must return to its initial position after each cycle.

A potential mechanism for the drift away from the wall may be the following.
During the half cycle of closer proximity of the particle to the wall, strains in the surrounding material reach higher values. The material has to support the same forcing in a more restricted and narrow geometry. 
Due to these higher strains of the surrounding material, elastic strains take longer to relax than those accumulated while the particle is further away from the wall. The larger stored elastic strain is thus able to push the particle further away from the wall with each cycle, leading to the observed drift.

The higher-order Fourier coefficients $\gamma_n$ provide further insight into the nonlinear response. 
In particular, a nonzero $\gamma_2$ indicates an asymmetric stress-strain curve, which arises due to one wall being closer than the opposite one.
The relative amplitude of the second harmonic of strain $|\gamma_2|/|\gamma_1|$ is shown in Fig.~\ref{fig:gamma-n_off-center}(b), marked by inverted triangles. 
It initially decreases with increasing $\omega\tau$, exhibits a shallow minimum around $\omega\tau \approx 1$, and subsequently increases, before saturating in the viscoelastic solid-like case of $\omega\tau \gg1$. 
The relative amplitude of the  third harmonic of strain $|\gamma_3|/|\gamma_1|$, marked in Fig.~\ref{fig:gamma-n_off-center}(b) by pentagons, shows a similar trend.
It decreases when turning from the viscous fluid-like to the viscoelastic regime and exhibits a local minimum at $\omega\tau \approx 1$.
However, instead of saturating in the solid-like limit, $|\gamma_3|/|\gamma_1|$ decreases to almost zero. The overall trend of $|\gamma_3|/|\gamma_1|$ is similar to that for the symmetric system, see Fig.~\ref{fig:gamma-n_center}.

\subsection{Microrheological response of two probe particles}
Finally, instead of a single probe particle, we consider a pair of probe particles subjected to oscillatory driving forces that are equal in magnitude but opposite in direction \cite{puljiz2018reversible}. The force acting on the $i$-th particle is given by
\begin{equation}
    \boldsymbol{F}_{1,2}(t) = {}\pm F(t) \hat{\boldsymbol{e}}_{x},
\end{equation}
where $F(t) = F_0\cos\omega t$.
Concerning the fluid particle dynamics approach, we now have to consider two phase fields $\phi_{1,2}(\boldsymbol{r})$ characterizing the particles.
Accordingly, the force density entering the Navier--Stokes equation is given as
\begin{equation}
\boldsymbol{f} = \frac{1}{V_\mathrm{p}}\left(\phi_1 \boldsymbol{F}_1  + \phi_2 \boldsymbol{F}_2 \right) .
\label{eq:TP_force-density}
\end{equation}
Denoting by $\Delta R(t) = |\boldsymbol{R}_1(t) - \boldsymbol{R}_2(t)|$ the instantaneous separation between the two probe particles, we define the strain as 
\begin{equation}
    \gamma(t) = \frac{\Delta R(t) - \Delta R(0)}{\Delta R(0)}.
\end{equation}
Similarly, the stress becomes
\begin{equation}
    s(t) = {}-\frac{F(t)}{L_{\mathrm{eff}}^{d-1}},
\end{equation}
where we now use the initial particle center-to-center distance as the effective length scale, $L_{\mathrm{eff}} = \Delta R(0)$.
The two-particle setup is considered in a periodic system without walls.
\begin{figure}
    \centering
    \includegraphics[width=\linewidth]{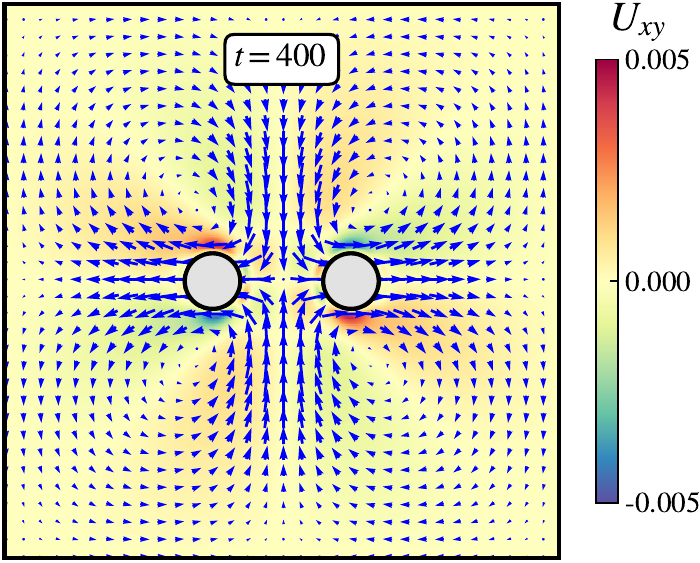}
    \caption{Snapshot of the two-particle simulation setup at time $t = 400$. The domain size is $20a\times 20a$ in terms of the particle radius $a$. The centers of the probe particles are initially $5a$ apart. Blue arrows indicate the velocity field, and the colormap quantifies the component  $U_{xy}$ of the strain tensor $\boldsymbol{U}$. Other parameters after rescaling are set to $\tau = 100$, $\mu_0 = 0.05$, $F_0 = 0.01$, and $\omega = 0.002\pi$.}
    \label{fig:TP-snapshot}
\end{figure}
Thus, here, the dynamics in the nonlinear regime of one particle is influenced by another particle, instead of a rigid boundary.
We set the initial particle center-to-center distance to $5a$.
Figure~\ref{fig:TP-snapshot} shows a snapshot of the system.

\begin{figure}
\centering
\includegraphics[width=\linewidth]{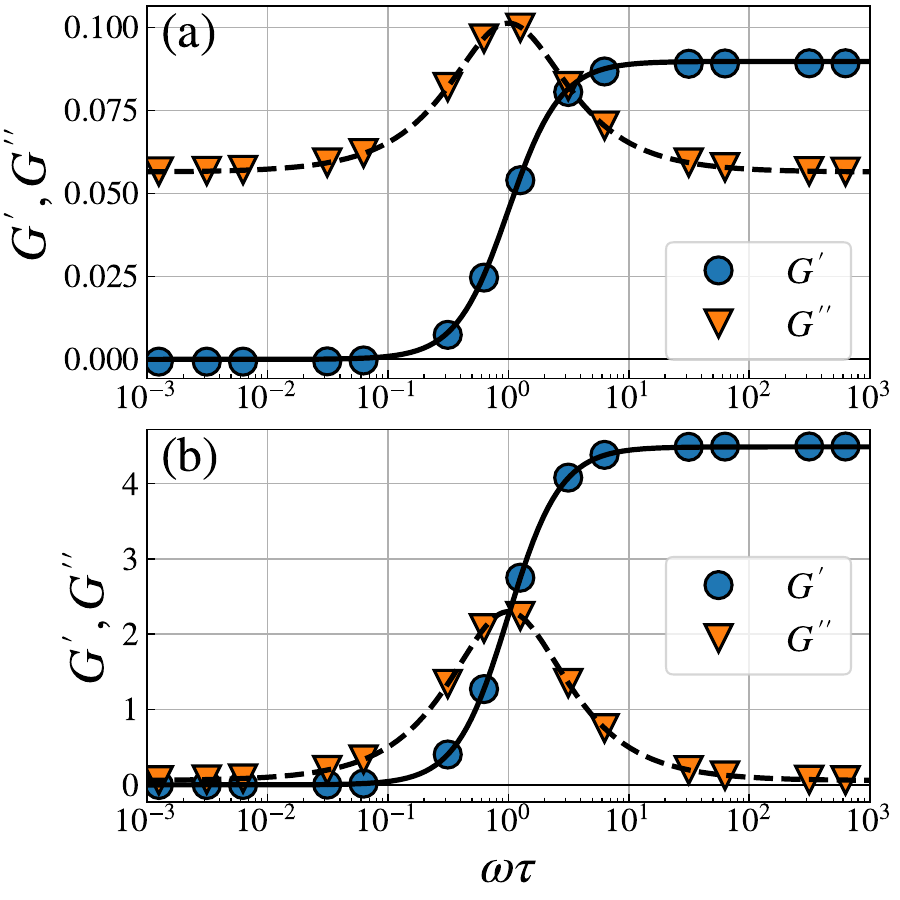}
\caption{For the two-particle simulation setup in the linear regime, the storage modulus $G'$ and the loss modulus $G''$ are plotted as functions of the Deborah number $\omega \tau$ for (a) $\mu_0 = 0.01$ and (b) $\mu_0 = 0.5$. The centers of the probe particles are initially a distance of $5a$ apart. Solid and dashed lines represent fits to $G'$ and $G''$, respectively, using Eqs.~\eqref{eq:storage-modulus} and \eqref{eq:loss-modulus}.
The only fit parameter is again the geometric prefactor, here obtained as $\alpha \approx 8.97$.
We fix the amplitude of the rescaled oscillatory driving forces at $F_0 = 0.01$.}
\label{fig:TP-linear-moduli}
\end{figure}

\begin{figure*}
\centering
\includegraphics[width=\linewidth]{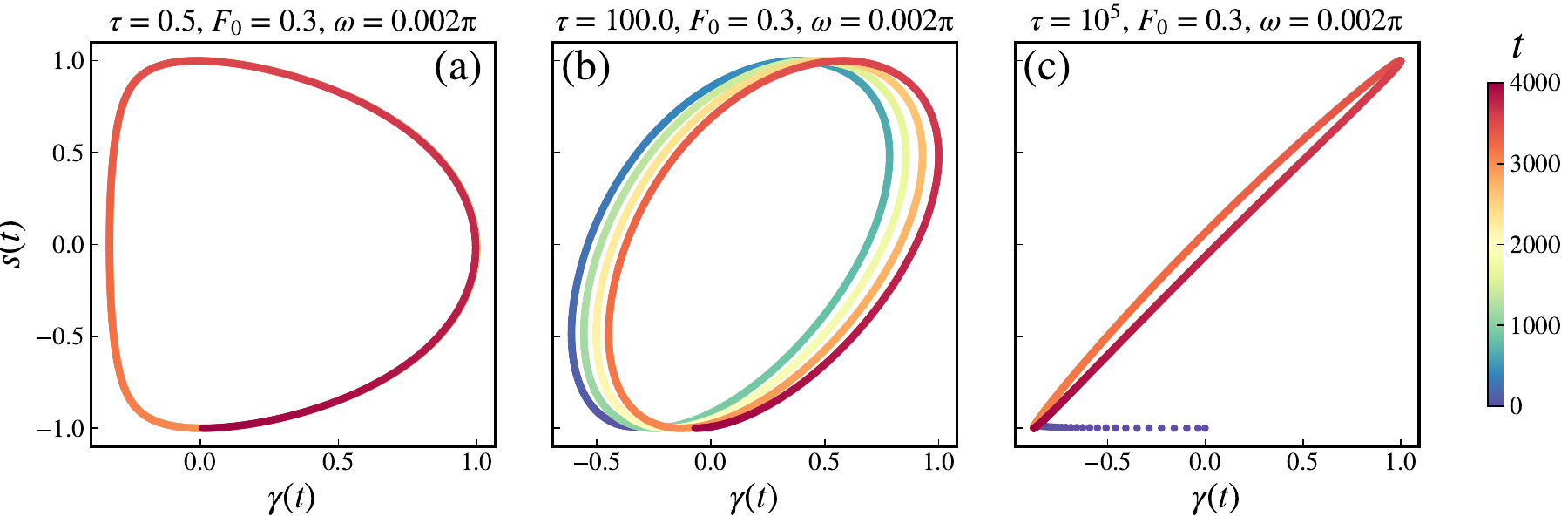}
\caption{Microrheological stress-strain curves in the nonlinear regime, but now for the two-particle setup at the three different values of the viscoelastic relaxation time (a) $\tau = 0.5$, (b) $\tau = 100$, and (c) $\tau = 10^5$.  
Again, the effects of increasing elastic contributions from left to right are visible. Moreover, nonlinearity and asymmetry of the setup when viewed from the perspective of one of the two particles become visible. Initially, the centers of the probe particles are $5a$ apart. 
The curves cover the first four oscillation cycles, with the color scheme indicating time $t$.
Remaining rescaled parameters are $F_0 = 0.3$, $\mu_0 = 0.1$, and $\omega = 0.002\pi$.}
\label{fig:TP-stress-strain-nonlinear}
\end{figure*}

\begin{figure}
\centering
\includegraphics[width=\linewidth]{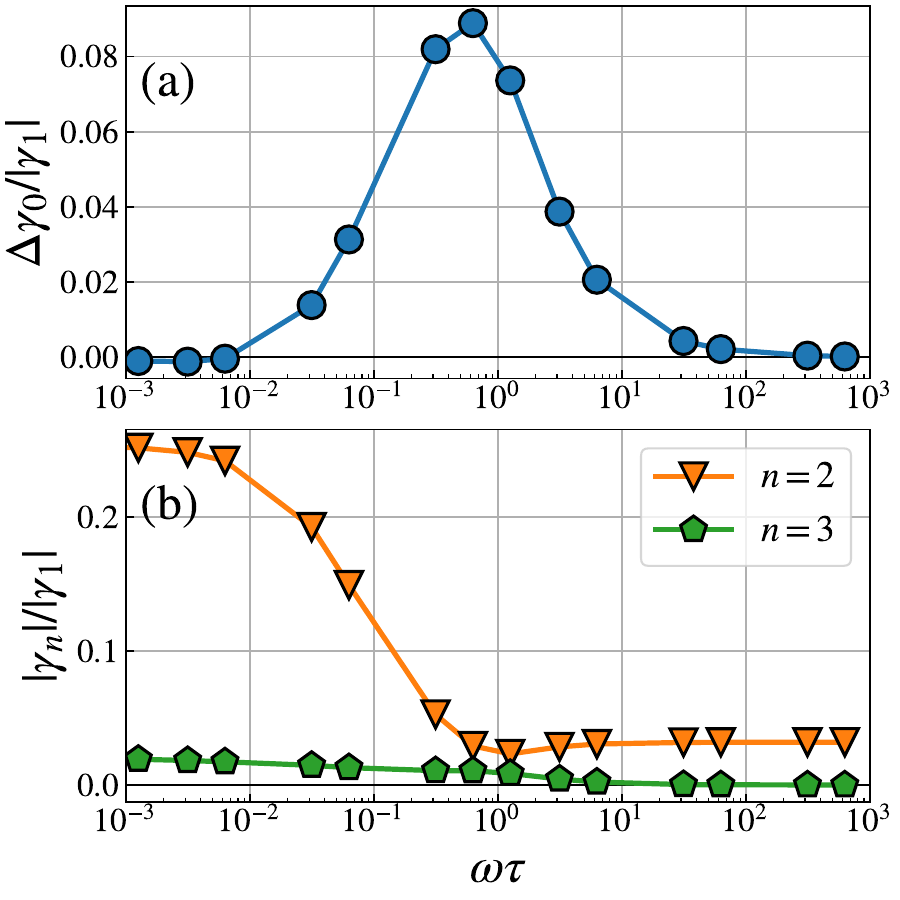}
\caption{Coefficients of the zero, second, and third harmonics of the induced strain, see Eq.~\eqref{eq:fourier-series}, for the microrheological two-particle setup in the nonlinear regime, see Fig.~\ref{fig:TP-stress-strain-nonlinear}, as a function of the Deborah number $\omega\tau$. Initially, the centers of the probe particles are $5a$ apart. (a) Relative difference $\Delta\gamma_0/|\gamma_1|$ in the values of the zero mode between the second and third cycle to quantify the associated drift in distance between the particles. (b) The relative amplitude of the second harmonic $|\gamma_2|/|\gamma_1|$ indicates the asymmetry in the underlying microrheological setup when viewed from the perspective of one particle. Besides, the relative amplitude of the third harmonic $|\gamma_3|/|\gamma_1|$ shows a trend similar to the one-particle case in Fig.~\ref{fig:gamma-n_off-center}. Again, we take averages over the second and third cycle. Remaining fixed rescaled parameter values are $F_0 = 0.3$, $\mu_0 = 0.1$, and $\omega = 0.002\pi$.}
\label{fig:TP-gamma-n}
\end{figure}

Similarly to the one-particle setup, we first focus on small amplitudes of the driving forces, here $F_0 = 0.01$, resulting in linear response.
To determine effective storage and loss moduli, $G'$ and $G''$, we again reproduce the curves expected from the linear Jeffreys model when varying the Deborah number $\omega \tau$.
Results are shown in Fig.~\ref{fig:TP-linear-moduli} for  $\mu_0 = 0.01$ and $\mu_0 = 0.5$. 
Here, the geometric fit parameter is obtained as $\alpha \approx 8.97$.

Increasing the amplitude of the oscillatory driving forces to $F_0 = 0.3$, nonlinear effects become visible.
Figure~\ref{fig:TP-stress-strain-nonlinear} shows the corresponding stress-strain curves, again for three values of the relaxation time.
The viscous fluid-like case for $\tau = 0.5$ is displayed in Fig.~\ref{fig:TP-stress-strain-nonlinear}(a) and shows a strong distortion of the stress-strain curve when comparing to the circular curve observed in the linear case in Fig.~\ref{fig:stress-strain-linear}(a).
Figure~\ref{fig:TP-stress-strain-nonlinear}(b) shows characteristic behavior in the viscoelastic fluid-like regime at $\tau = 100$.
Here, we again observe a net drift of the particle positions.
Similarly to the one-particle setup, where the particle moves successively further away from the closer wall, the two probe particles move further apart from each other with every successive oscillation cycle.
Finally, Fig.~\ref{fig:TP-stress-strain-nonlinear}(c) illustrates the behavior when we approach the solid-like limit.

To further illustrate these effects, we display the behavior of the coefficients of the zeroth, second, and third harmonic, see Eq.~\eqref{eq:fourier-series}, in Fig.~\ref{fig:TP-gamma-n}.
Mostly, their trends are similar to those of the nonsymmetric one-particle setup in close vicinity to a wall as shown in Fig.~\ref{fig:gamma-n_off-center}.
The normalized cycle-averaged drift $\Delta \gamma_0/|\gamma_1|$ displays a maximum when the inverse of the oscillation frequency approximately matches the relaxation time scale, $\omega\tau\approx1$.
The relative amplitude of the second harmonic $|\gamma_2|/|\gamma_1|$ shows an overall decreasing trend with increasing Deborah number $\omega \tau$. It exhibits a shallow local minimum at $\omega \tau \approx 1$.
Moreover, the relative amplitude of the third harmonic $|\gamma_3|/|\gamma_1|$ remains rather small. Yet, it shows an overall decreasing trend with increasing $\omega \tau$, as for the one-particle setup in Fig.~\ref{fig:gamma-n_off-center}.

These results for the two-particle system are in line with our previous observations on the bounded one-particle system.
In particular, they show that a viscoelastic cycle-averaged drift can likewise emerge as a consequence of particle--particle interactions.

\section{CONCLUSIONS}\label{sec:summary}
Summarizing, we investigated by computational means a setup of active microrheology in the linear and nonlinear regimes for viscous fluid-like, viscoelastic fluid-like, and viscoelastic quasi-solid systems. Our description allows to tune between these different systems by adjusting one viscoelastic relaxation time scale. To infer the microrheological behavior of the system, we evaluated the response of one or two probe particle(s) embedded in the surrounding medium to imposed oscillatory forces. In order to describe the coupling between the particle(s) and the surrounding material by computational means, we extended the fluid particle dynamics (FPD) method~\cite{tanaka2000simulation,tanaka2006viscoelastic} to viscoelastic media. Here, we worked with the neo-Hookean hyperelastic model, but other descriptions of nonlinear elasticity can likewise be employed.  

In the linear regime, the response of the system is described by a three-parameter Jeffreys model. The storage and loss mechanical moduli obtained from the simulations agree with the analytical predictions of this linear model. We recover the expected viscous and elastic limits for small and large Deborah numbers, respectively.

Beyond the linear regime, nonlinear effects emerge that cannot be described by a single complex mechanical modulus. 
Nonlinearities generate higher harmonics and, for nonsymmetric systems, a net drift.
We observe the latter when we place a probe particle off center into our system bounded by rigid walls. In this case, for viscoelastic fluids, we find a gradual drift of the mean particle position away from the closer wall over successive oscillation cycles.
We encounter an analogous effect in a two-particle setup, where particles successively move apart from each other with each cycle.
This drift effect 
is most pronounced when the viscoelastic relaxation timescale is close to the oscillation timescale of the imposed force.
It vanishes for viscous fluids and in the limit of elastic solids.
In a different context, particle drift and migration effects due to viscoelasticity are utilized in microfluidic devices~\cite{gaetano2017particle}.
However, in contrast to most microfluidic systems, the force in active microrheology is acting on the particles directly.

We recall that, in this initial extension to viscoelastic systems, we confined ourselves to planar, effectively two-dimensional setups. 
Various systems and geometries are described by such planar arrangements. They range from studies on biological membranes \cite{danelon2006cell}, via monolayers of graphene \cite{lee2008measurement}, to synthetic polymeric composite membranes \cite{jiang2004freely}. 
One way of investigating their properties is to span them over holes and analyze their mechanical response \cite{steltenkamp2006mechanical,lee2013high}. Further example systems are smectic liquid-crystalline films \cite{eremin2011two} or sheets of smectic liquid-crystalline elastomers in homeotropic alignment \cite{nishikawa1997smectic}, which tend to maintain their thickness to first approximation. Theoretically, plane-stress and plane-strain geometries can be mapped to two dimensions \cite{zisiadis2026effective}. Interactions between a particle and a rigid wall \cite{lutz2022effect} as well as between two particles experiencing pairwise reciprocal forces \cite{richter2022mediated} have been worked out by analytical theory in two dimensions in the (quasi)static elastic limit. Likewise, the geometry of a purely elastic, rigidly bounded, square-shaped domain has been addressed \cite{sprenger2024thin}. Our present work provides the extension of such studies to the viscoelastic, dynamic regime. 
In the future, a next step concerns extension to three dimensions, which is conceptually straightforward, although computationally significantly more intensive.

Here, we focused on a setup of active microrheology in terms of the translational motion of probe particles.
However, rotational effects in one- or multiple-particle systems can be utilized for microrheological probing as well~\cite{wilhelm2003rotational,bishop2004optical,schmiedeberg2005one,puljiz2016forces,richter2021rotating}.
Our extended FPD method may be applied to these setups in future studies.
Similarly, the simulation strategy can be readily adopted to large many-particle systems~\cite{tanaka2000simulation,tanaka2006viscoelastic,reinken2026hydrodynamics}.
In combination with our viscoelastic extension, such many-particle systems, driven by imposed or induced forces on the particles, can be studied in carrier media of more complex rheology.
Examples are magnetic or magnetically induced driving forces in the context of magnetorheological fluids~\cite{bossis2002magnetorheological,vicente2011magnetorheological} and elastomers~\cite{odenbach2016microstructure,bastola2020recent,fischer2026magnetic}. 
In these materials, the structuring of large numbers of magnetized micron-sized particles in response to an external magnetic field allows the external control of effective mechanical properties~\cite{holm2005structure,metsch2016numerical,fischer2019magnetostriction,fischer2024maximized,fischer2026doubling}.
Self-propelled microswimmers, such as bacteria moving in viscoelastic media~\cite{hemingway2015active,mathijssen2016upstream,puljiz2019memory,plan2020active,li2021microswimming,reinken2025unified,reinken2025rheologically}, represent another interesting topic for future studies.

\section*{Acknowledgements}
The authors thank the Deutsche Forschungsgemeinschaft (German Research Foundation, DFG) for support through Research Unit FOR 5599 on structured magnetic elastomers, project project nos.
511114185, 535421963, and 535543971 (DFG ref. nos. ME 3571/10-1 and ME 3571/11-1).

\section*{Author Declaractions}

\subsection*{Conflict of Interest}
The authors have no conflicts to disclose.

\subsection*{Author Contributions}

\textbf{Muhammed Muhsin Abdul Azeez}: Data Curation (lead), Formal Analysis (lead), Investigation (lead), Methodology (equal), Software (lead), Validation (equal), Writing: Original Draft Preparation (lead), Writing: Review \& Editing (equal). \textbf{Henning Reinken}: Conceptualization (supporting), Methodology (equal), Project Administration (equal), Software (supporting), Supervision (equal), Validation (equal), Writing: Original Draft Preparation (supporting), Writing: Review \& Editing (equal). \textbf{Andreas M. Menzel}: Conceptualization (lead), Funding Acquisition (lead), Methodology (equal), Project Administration (equal), Resources (lead), Supervision (equal), Writing: Review \& Editing (equal).

\section*{Data Availability}

The data that support the findings of this study are available from the corresponding author upon reasonable request. Data underlying the figures will be made publicly available on Zenodo after review. 

\appendix

\section{Rescaling}\label{App:rescaling}
To rescale the equations, we use the radius of the particle(s) $a$ as a length scale and the time $\rho a^2/\eta_0$ as a time scale. The corresponding dimensionless variables are defined as
\begin{equation}
    \boldsymbol{r} = a \tilde{\boldsymbol{r}}, \quad  \quad
    t = \frac{\rho a^2}{\eta_0} \tilde{t},
\end{equation}
where the tilde marks rescaled units. Thus, Eq.~\eqref{eq:navier-stokes} can be expressed in terms of the dimensionless variables as
\begin{equation}
    \partial_{\tilde{t}} \tilde{\boldsymbol{v}} + \tilde{\boldsymbol{v}}\cdot \tilde{\nabla} \tilde{\boldsymbol{v}} = \tilde{\nabla} \cdot \tilde{\boldsymbol{\sigma}} + \tilde{\boldsymbol{f}}_\mathrm{p} + \tilde{\boldsymbol{f}}_\mathrm{w}.
    \label{eq:navier-stokes-rescaled}
\end{equation}
Here, 
\begin{equation}
    \tilde{\boldsymbol{\sigma}} = - \tilde{p} \boldsymbol{I} +  \tilde{\eta} \left[ \tilde{\nabla} \tilde{\boldsymbol{v}} + (\tilde{\nabla} \tilde{\boldsymbol{v}})^\mathsf{T} \right]  + \tilde{\mu} \boldsymbol{B},
\end{equation}
$\eta=\eta_0\tilde{\eta}$, 
$\tilde{\eta}(\tilde{\boldsymbol{r}}) = 1 + (R_{\eta} - 1)\phi
(\tilde{\boldsymbol{r}})$, and ${\mu} = ({\eta_0^2}/{\rho a^2}) \tilde{\mu}$. Moreover, $p= ({\eta_0^2}/{\rho a^2})\tilde{p}$. 
Likewise, the rescaled force densities are given by ${\boldsymbol{f}}_{\mathrm{p}/\mathrm{w}} = ({\eta_0^2}/{\rho a^3})\tilde{\boldsymbol{f}}_{\mathrm{p}/\mathrm{w}}$. Similarly, the dynamic equation for $\boldsymbol{U}$,  Eq.~\eqref{eq:U-evolution}, becomes
\begin{equation}
    \partial_{\tilde{t}} \boldsymbol{U} + (\tilde{\boldsymbol{v}} \cdot \tilde{\nabla})\boldsymbol{U} + (\tilde{\nabla} \tilde{\boldsymbol{v}})\cdot \boldsymbol{U} + \boldsymbol{U}\cdot (\tilde{\nabla} \tilde{\boldsymbol{v}})^{\mathsf{T}}= \tilde{\boldsymbol{A}}  - \frac{1}{\tilde{\tau}} \boldsymbol{U},
    \label{eq:U-evolution-rescaled}
\end{equation}
where ${\tau} = (\rho a^2/\eta_0)\tilde{\tau}$. 

The dimensionless equations presented above are the ones used throughout the numerical simulations. For simplicity, tildes are mostly omitted in the main text.

\section{Details on the Simulation Methods}\label{App:numerical-methods}
To simulate the system, we couple the discrete particle description with the continuum hydrodynamic description of the surrounding medium~\cite{reinken2026hydrodynamics}. This is achieved by representing the particle through a phase field $\phi$, which distinguishes the particle region from the surrounding medium. For every time step, we calculate the force $\boldsymbol{F}(t)$ acting on the particle(s). This force is then incorporated into the Navier--Stokes equation, Eq.~\eqref{eq:force-density}, through the force density $\boldsymbol{f}_\mathrm{p}$. 

Equations~\eqref{eq:navier-stokes-rescaled} and \eqref{eq:U-evolution-rescaled} are discretized in time using the implicit Euler method. 
We employ an iteration procedure during each time step to solve the implicit equation and find the solution at each new time step.
Here, the spatial derivatives are calculated in Fourier space based on a pseudo-spectral approach~\cite{canuto2007spectral}. 
To improve the convergence and stability of the iteration scheme, a relaxation procedure is employed.
In each step of iteration $k$ of finding the solution at the next time step, the updated velocity and strain tensor fields, $\boldsymbol{v}^{k+1}$ and $\boldsymbol{U}^{k+1}$, are obtained by mixing the newly estimated solutions, $\boldsymbol{v}_\mathrm{e}$ and $\boldsymbol{U}_\mathrm{e}$, with those of the current iteration step, $\boldsymbol{v}^{k}$ and $\boldsymbol{U}^{k}$.
Here, $\boldsymbol{v}_\mathrm{e}$ and $\boldsymbol{U}_\mathrm{e}$ are estimated based on the velocity and strain tensor fields in the current iteration step, $\boldsymbol{v}^k$ and $\boldsymbol{U}^k$.
In every iteration step, we thus obtain $\boldsymbol{v}^{k+1}$ and  $\boldsymbol{U}^{k+1}$via 
\begin{equation}
\begin{aligned}
    \boldsymbol{v}^{k+1} &\leftarrow (1 - w)\boldsymbol{v}^{k} + w \ \boldsymbol{v}_\mathrm{e}(\boldsymbol{v}^{k},\boldsymbol{U}^k),\\
    \boldsymbol{U}^{k+1} &\leftarrow (1 - w)\boldsymbol{U}^{k} + w \ \boldsymbol{U}_\mathrm{e}(\boldsymbol{v}^k,\boldsymbol{U}^{k}),
    \end{aligned}
\end{equation}
The weight parameter $w$ controls the relaxation strength. $0~<~w~<~1$ corresponds to underrelaxation, whereas $w>1$ corresponds to overrelaxation. 
In the present work, we adaptively change $w$ between the bounds $0.1 \le w \le 1.5$ depending on how the remaining errors have changed during the iteration step. 
Here, the relative errors in the velocity and strain tensor fields are defined as 
\begin{equation}
\begin{aligned}
\varepsilon_{\boldsymbol{v}} &= \frac{|\boldsymbol{v}_{\mathrm{e}}(\boldsymbol{v}^k,\boldsymbol{U}^{k}) - \boldsymbol{v}^k|}{\sqrt{\langle |\boldsymbol{v}^k|^2\rangle}},\\
\varepsilon_{\boldsymbol{U}} &= \frac{||\boldsymbol{U}_\mathrm{e}(\boldsymbol{v}^k,\boldsymbol{U}^{k}) - \boldsymbol{U}^k||_\mathrm{F}}{\sqrt{\langle || \boldsymbol{U}^k||_\mathrm{F}^2\rangle}},
\end{aligned}
\end{equation}
where $|| \dots ||_\mathrm{F}$ denotes the Frobenius norm of a tensor and $\langle \dots\rangle$ is the spatial average.
The value of $w$ is adaptively increased when both the maximum local errors, $\mathrm{max}(\varepsilon_{\boldsymbol{v}})$ and  $\mathrm{max}(\varepsilon_{\boldsymbol{U}})$, have decreased during the previous iteration step.
Otherwise, $w$ is lowered.
This procedure is repeated until the maximum local errors are both less than $0.1\% $. The pressure $\tilde{p}$ in Eq.~\eqref{eq:navier-stokes-rescaled} is determined by enforcing incompressibility at each iteration step via a projection method~\cite{durran2010numerical}. 

We incorporate the effect of rigid, immovable walls at the boundaries of the system via the additional force density $\boldsymbol{f}_\mathrm{w}$.
Here, we use finitely extended walls of thickness $a_\mathrm{w}$.
Similar to the field $\phi$ marking the presence of the particle(s), the walls are represented by a phase field $\phi_\mathrm{w}$, 
\begin{equation}
\phi_\mathrm{w}(\boldsymbol{r}) = \frac{1}{2}\left[\tanh{\left(\frac{a_\mathrm{w} - |\boldsymbol{r} - \boldsymbol{r}_\mathrm{b}|}{c_\mathrm{w}} \right)} + 1\right],
\label{eq:phiFieldWall}
\end{equation}
where $\boldsymbol{r}_\mathrm{b}$ is the point on the boundary of the calculation box closest to the position $\boldsymbol{r}$ and $c_\mathrm{w}$ sets the thickness of the numerical interface between the wall and the enclosed medium.
Assuming no-slip boundary conditions between the viscoelastic medium and the rigid, immovable walls, the purpose of the force density $\boldsymbol{f}_\mathrm{w}$ is to ensure that the velocity $\boldsymbol{v}$ remains zero within the area described by the field $\phi_\mathrm{w}$.
To this end, we employ a strong linear frictional damping scheme~\cite{goldstein1993modeling,kevlahan2001computational}. 
The force density in this framework is calculated as
\begin{equation}
\boldsymbol{f}_\mathrm{w} = - \phi_\mathrm{w} \beta \boldsymbol{v},
\label{eq:wallForcing}
\end{equation}
where $\beta$ sets the strength of the damping.
For our numerical simulations, we use $\beta = 10^3$.
Moreover, we set the thickness of the walls equal to the radius of the particle(s), $a_\mathrm{w} = a$. We use the same thickness of the interface region as for the particle(s), $c_\mathrm{w} = c$.

Having obtained the velocity field $\boldsymbol{v}$ from the continuum description, we can turn to the particle dynamics. The particle velocity $\boldsymbol{V}(t)$ is obtained by integrating $\boldsymbol{v}(\boldsymbol{r},t)$ over the particle body, weighted by the phase field $\phi(\boldsymbol{r},t)$,
\begin{equation}
    \boldsymbol{V}(t) = \frac{1}{V_\mathrm{p}}\int \mathrm{d}\boldsymbol{r}\; \boldsymbol{v}(\boldsymbol{r},t) \phi(\boldsymbol{r},t).
\end{equation}
The particle position $\boldsymbol{R}(t)$ is then updated using an explicit Euler method,
\begin{equation}
    \boldsymbol{R}(t + \Delta t) = \boldsymbol{R}(t) + \boldsymbol{V}(t)\Delta t.
\end{equation}
Finally, the phase field $\phi$ is updated according to Eq.~\eqref{eq:phi-filed}, using the updated particle position.

\bibliography{references}

\end{document}